\documentclass[epj,english,final]{svjourarX}
\usepackage[numbers,sort&compress]{natbib}
\usepackage[T1]{fontenc}
\usepackage[latin9]{inputenc}
\usepackage{amsmath}
\usepackage{graphicx}

\makeatletter
 \PassOptionsToPackage{caption=false}{subfig}
\usepackage{subfig}

\makeatother

\usepackage{babel}

\begin{document}
\global\long\def\dd{\,\mathrm{d}}

\global\long\def\ii{\mathrm{i}}

\global\long\def\ee{\mathrm{e}}

\global\long\def\Sp{\mathrm{Sp}}

\global\long\def\lb{\mathrm{lb}}

\global\long\def\bra#1{\left<#1\right|}

\global\long\def\ket#1{\left|#1\right>}

\global\long\def\braket#1#2{\left<#1\vphantom{#2}\right|\left.\hspace{-0.25em}#2\vphantom{#1}\right>}

\global\long\def\ketbra#1#2{\left|#1\vphantom{#2}\right>\hspace{-0.4em}\left<#2\vphantom{#1}\right|}

\global\long\def\esti{^{\mathrm{est}}}

\global\long\def\expe{^{\mathrm{exp}}}

\global\long\def\absq#1{\left|#1\right|^{2}}

%\title{Simple Adaption of Measurements for Qu{\sf\bf\it d}it Estimation}
\title{Simple Adaption of Measurements for Qu$d$it Estimation}

\author{Christof J. Happ \and Matthias Freyberger}

\institute{Institut f\"ur Quantenphysik, Universit\"at Ulm, D-89069 Ulm, Germany}

%\date{Received: 15 June 2011 / Revised version: date}
\date{ }

\abstract{
We present a strategy for estimation of $d$-level quantum states
and for the simple adaption of corresponding measurements.
The adaption method is inspired by mutually unbiased measurements,
but it is also applicable in cases for which no complete set of mutually
unbiased bases is known. We present results of Monte Carlo simulations,
that demonstrate the fidelity gain of the adaptive strategy compared
to a non-adaptive estimation.
% end of PACS codes
}%end of abstract
\maketitle

\section{Introduction}

From the viewpoint of fundamental quantum mechanics 
as well as from the viewpoint of practical applications in quantum information science, 
determining the state of a quantum system \cite{ParisRehacek,Raymer97} is an important problem,
very different from the measuring process in classical physics. 
Perfect reconstruction of a completely unknown quantum state requires an infinite
amount of measurements for various non-commuting observables. 
One thus also needs an infinite number of identical
quantum systems. However, due to the no-cloning theorem \cite{Dieks82,WZ82}
they cannot be copies of a single system. As a consequence, initial copies,
or more exactly, identically prepared systems, are a valuable resource.
From this limited resource only a limited amount of information can
be extracted.

This contribution aims at an optimised utilisation of this resource for the estimation
of a pure $d$-level quantum state by single measurements. The benchmark for
the quality of such a procedure is the optimal average fidelity of
a joint measurement \cite{MP95,DBE98,BM99} on all systems of the finite
sample. However, the needed physical principles for such
collective measurements are easily realised. Hence we will 
analyse how well this optimal method can be approximated by an approach
using measurements on single copies and  an adaptive choice of measurement
parameters.

For qubits such single-copy measurements have been described for pure
\cite{HSBR00,BBMt02,BMMt05} and mixed states \cite{BBGMMt06}. An
adaptive method has been reported in \cite{FKF00} for qubits as well.
There it was possible to come quite close to the theoretical limit given
in \cite{MP95}. For higher dimensional states single measurement
schemes based on maximum likelihood estimation \cite{Hradil97,ParisRehacek}
and mutually unbiased measurements \cite{WF89} have already been used in
experiments \cite{KMFS08,AS10,LimaEtAl11}. However the application of
adaptive methods to higher dimensions is not straightforward. The
methods used for estimation and adaption cannot be easily calculated
in general, and the numerical expense to simulate these
methods efficiently turns out to be quite high. 
Therefore, alternative and  simpler methods are needed. 
Before we explain these in detail, let us recall the basic problem.

\section{Basic Concepts}

A finite number of identically prepared versions of a $d$-dimensional
quantum system shall be given. We have no prior information
on the corresponding pure state $\ket{\psi}$ and 
want to infer it from the results of single measurements.
To this end, on one hand we have to reasonably estimate a state $\ket{\psi\esti}$
from already gathered data (estimation). On the other hand we have
to decide which measurement parameters to use for further measurement
steps (adaption). We restrict ourselves to projective measurements,
which can be realised in principle.
Hence the main question is for the appropriate choice 
of the measurement basis $\left\{ \ket{b_{i}}\right\} $. 
This choice can be influenced by previously accumulated
measurement data only. Therefore, in the beginning we can use an arbitrary
basis, e.g. the computational basis. But after the first measurement
we can adapt the next basis according to the previous results, conduct
this measurement, and continue as long as there are copies left.

In the end, we have acquired a sequence of used measurement bases
and the corresponding results, i.\,e. which vector of each basis
was really found in the measurement. Based on this measurement data
we perform the estimation procedure, which gives us the estimated
state $\ket{\psi\esti}$. To assess the quality of estimation, we
can use the fidelity $F=\left|\braket{\psi}{\psi\esti}\right|^{2}$.
In general, it will of course be different for different unknown states
$\ket{\psi}$. But more importantly, due to the probabilistic nature
of quantum measurements the fidelity even differs for different measurement
runs with the same state $\ket{\psi}$. While we expect from a good
estimator to work equally well for all unknown states, the dependence
on the measurement results is inevitable. Therefore, to arrive at
a valid criterion for characterising the adaptive procedure, we have to
average over all possible results. In practise we achieve this by
averaging over sufficiently many Monte Carlo simulations of such measurement
runs.

\section{Estimation}

The key part of state estimation is transforming the measurement data
into an estimate of the unknown quantum state. A useful method for qubits
is maximum likelihood estimation \cite{Hradil97,ParisRehacek}. But
due to the growing number of state parameters in higher
dimensions sooner or later this becomes highly involved. This is equally well a
problem for an experimenter, who has to find an estimate for an unknown
state, as for a theoretician, who wants to assess numerically
the quality of a measurement scheme. The latter has to cope with the
additional complication that one has to average over a statistically
significant number of simulated measurement runs. 

Therefore we propose another much more primitive estimation method.
For qubits this means calculating the average of the Bloch vectors corresponding
to the measured states and using this average as representation of
a (mixed) estimated state.%
\footnote{Note that this method gives unbiased information on the direction
of the Bloch vector only, while its length is not only determined
by the purity of the unknown state but also by the distribution of the
measured states.%
} In order to get a pure estimated state, this Bloch vector can be
normalised. This can be viewed as a very rough simplification of maximum
likelihood estimation, because the likelihood function for a single
measurement is centred at the measured point on the Bloch sphere and
the likelihood for more measurements is the product of such elementary
likelihoods. Since a Bloch vector is nothing more than
a representation of the density matrix,
this is the same as directly calculating
the average density matrix and choosing the pure state nearest to
this average matrix.

This simple averaging procedure can be generalised to higher dimensions $d$.
As in the qubit case a geometric representation of the state can be used.
One such generalisation of the Bloch vector, that allows this averaging procedure,
is the coherence vector \cite{HE81,MahlerWeberruss}.
On the other hand the average density matrix $\hat{\overline{\rho}}$ can be calculated directly:
We simply add all found measurement results $\left\{\ket{m_{k}}\right\}$ incoherently and
arrive at
\begin{equation}
\hat{\overline{\text{\ensuremath{\rho}}}}=\tfrac{1}{\nu}\sum_{k=1}^{\nu}\ketbra{m_{k}}{m_{k}} 
\end{equation}
after $\nu$ measurements.
However, in the subsequent search for a pure estimated state, a new complication
arises, compared to the $d=2$ case, where it was sufficient 
to normalise the Bloch vector of $\hat{\overline{\rho}}$.

In order to find the pure state in best agreement with the measured
data, we choose as estimated state $\ensuremath{\ket{\psi\esti}}$
the one with maximal overlap with the averaged density operator $\hat{\overline{\rho}}$.
For eigenvalues $e_{j}$ and eigenvectors $\ket{e_{j}}$ of $\hat{\overline{\rho}}$ we arrive at 
\begin{equation}
\bra{\psi\esti}\hat{\text{\ensuremath{\overline{\rho}}}}\ket{\psi\esti}=\sum_{j=0}^{d-1}e_{j}\absq{c_{j}}.
\end{equation}
with $c_{j}=\braket{e_{j}}{\psi\esti}$. 
When we arrange the eigenvalues in descending order $e_{0}\geq e_{1}\geq\ldots\geq e_{d-1}$,
we find 
\begin{equation}
\bra{\psi\esti}\hat{\text{\ensuremath{\overline{\rho}}}}\ket{\psi\esti}\leq e_{0},
\end{equation}
where equality holds only if $c_{0}=1$ and all other coefficients
vanish.

This means that the wanted estimate $\ket{\psi\esti}=\ket{e_{0}}$
is the eigenvector corresponding to the largest eigenvalue $e_{0}$
of $\hat{\overline{\rho}}$ (and the eigenvalue $e_{0}$ itself is
the overlap of $\hat{\overline{\rho}}$ and $\ensuremath{\ket{\psi\esti}}$).
Thus the estimation procedure consists of two numerically simple steps:
incoherent superposition of the measured states and solving the eigenvalue
problem of the averaged density matrix.
For a comparison of this estimation scheme 
to other possible estimators applied to non-adaptive measurement setups see \cite{BPsF06}.

\section{Least Bias Adaption}

We have described a method for calculating estimated states from measurement
data; but we still have to decide which measurement bases to use.
In the qubit case it was possible to formulate an expected fidelity
with respect to the used measurement parameters and to find the optimal
parameters by maximising this expected fidelity \cite{FKF00,HF08}.
Again this becomes a tedious numerical task in higher dimensions. Instead
we use a method that mimics mutually unbiased bases (MUBs) \cite{WF89}.

Two bases $\left\{ \ket{b_{i}}\right\} $ and $\left\{ \ket{b_{j}'}\right\} $
are called unbiased if each pair of basis vectors from different bases
has the same overlap
\begin{equation}\left|\braket{b_{i}}{b_{j}'}\right|^{2}=\tfrac{1}{d}.\end{equation}
In $d$ dimensions there are up to $d+1$ bases that are mutually unbiased
\cite{DEBZ10}. Such sets of MUBs are optimal
for state estimation with fixed measurements on single copies in the
sense of minimising statistical errors \cite{WF89,LimaEtAl11}.
However, as soon as there are more measurements performed than are MUBs known, 
it is possible to find measurement strategies that improve estimation
fidelity compared to repeated MUB measurements \cite{BBMt02}. Our
least bias adaption method broadens the idea of mutually unbiased
measurements for use with more measurements.

The ultimate goal of adaption is finding measurement parameters that
maximise the information gained in this measurement. Mutually unbiased
measurements ensure exactly that, because each sequence of possibly
measured state vectors is {}``unbiased'', having overlap $\tfrac{1}{d}$
with each other. However, since the measurements are made successively,
mutual unbiasedness is sufficient but not necessary for this maximised information gain. 
Only the actually measured basis vector has to be remembered from each measurement; 
the directions of the other basis vectors are irrelevant. Therefore the next measurement
basis has to be {}``unbiased'' with respect to the previous measurement
results but not to the previous measurement basis vectors which were
never actually found in a measurement. Using this relaxed definition
of unbiasedness, more measurement bases can be constructed. 
As another consequence of this concept of unbiasedness, 
it depends on the previous results which measurements can be called unbiased. 
Therefore, this is a truly adaptive method.
Of course, at some point Hilbert space is {}``filled'' and no more
unbiased bases can be found. Then the adaption method must
be changed to not choosing an unbiased basis, but rather the basis with least
bias with respect to the measured vectors $\left\{ \ket{m_{k}}\right\} $. 

For developing this idea into an adaption algorithm, we have to define
a function assessing bias in the above sense. It should yield unbiased
bases as long as this is still possible. A good choice for such
a function is the entropy-like function
\begin{equation}
h=-\sum_{j=0}^{d-1}\sum_{k=1}^{\nu}\left|\braket{m_{k}}{b_{j}}\right|^{2}
\ln\left|\braket{m_{k}}{b_{j}}\right|^{2},\label{eq:BiasCont}
\end{equation}
where $\left\{ \ket{m_{k}}\right\} $ are the $\nu$ vectors corresponding
to already found results of the $\nu$ preceding measurements and
$\left\{ \ket{b_{j}}\right\} $ is the orthogonal basis to be adapted. 

This can be interpreted as the conditional entropy 
$h=\sum_{j,k}p(m_{k})p(b_{j}|m_{k})\ln p(b_{j}|m_{k})$,
if we set all $p(m_{k})$ to be equally {}``true'' and thus equally
likely \cite{NielsenChuang,CoverThomas}. Hence $h$ quantifies the information gained when measuring
in $\left\{ \ket{b_{j}}\right\} $ if the {}``signals'' $\left\{ \ket{m_{k}}\right\} $
are already known. 

The basis maximising conditional entropy will be adap\-ted and chosen
for the next measurement. This will be done by the equal distribution
$\left|\braket{m_{k}}{b_{j}}\right|^{2}=\frac{1}{d}\;\forall k,j,$
if this is possible, i.\,e. there are not too many measurements, and
we thus find a basis unbiased with respect to the $\left\{ \ket{m_{k}}\right\} $.
At some point there will be too many terms in equation (\ref{eq:BiasCont})
to allow this kind of unbiasedness, but the adaption algorithm can
still choose the basis maximising $h$ for the next measurement.

In order to maximise the conditional entropy (\ref{eq:BiasCont}) correctly,
we have to ensure that the set $\left\{ \ket{b_{j}}\right\} $ is
a proper basis, i.\,e. the elements of $\left\{ \ket{b_{j}}\right\} $
are not arbitrary vectors, but have to fulfil the condition of orthonormality.
Since all bases can be generated from an arbitrary computational basis
by a unitary transformation, this is equivalent to parameterising
a unitary matrix. We did this using the Hurwitz construction \cite{Hurwitz97,ZK94},
producing such a unitary matrix from elementary two-dimensional rotations.

\section{Monte Carlo Simulations}

To demonstrate the influence of our least bias strategy on estimation fidelity
in different dimensions, we present Monte Carlo simulations of state
estimation procedures using least bias estimation and compare these
to a non-adaptive estimation scheme. 
The latter is realised by choosing the unitary matrix that rotates the computational basis randomly. 
In order to arrive at an equal distribution of the matrices, we drew their parameters according to the relevant Haar measure \cite{Hurwitz97,ZK94}.

Each estimation run uses up to 50 measurements
(and would therefore require the same amount of identically prepared systems). 
Since the probabilistic nature of quantum measurements heavily influences
the fidelity of each single run, we average over a statistically significant
amount of such estimation runs (typically $10^{4}$). 
This average fidelity is shown in the following figures with respect to the used copies.

\subsection{Qubits}
\begin{figure}[h]
\subfloat[fidelity]{\includegraphics[width=0.5\textwidth]{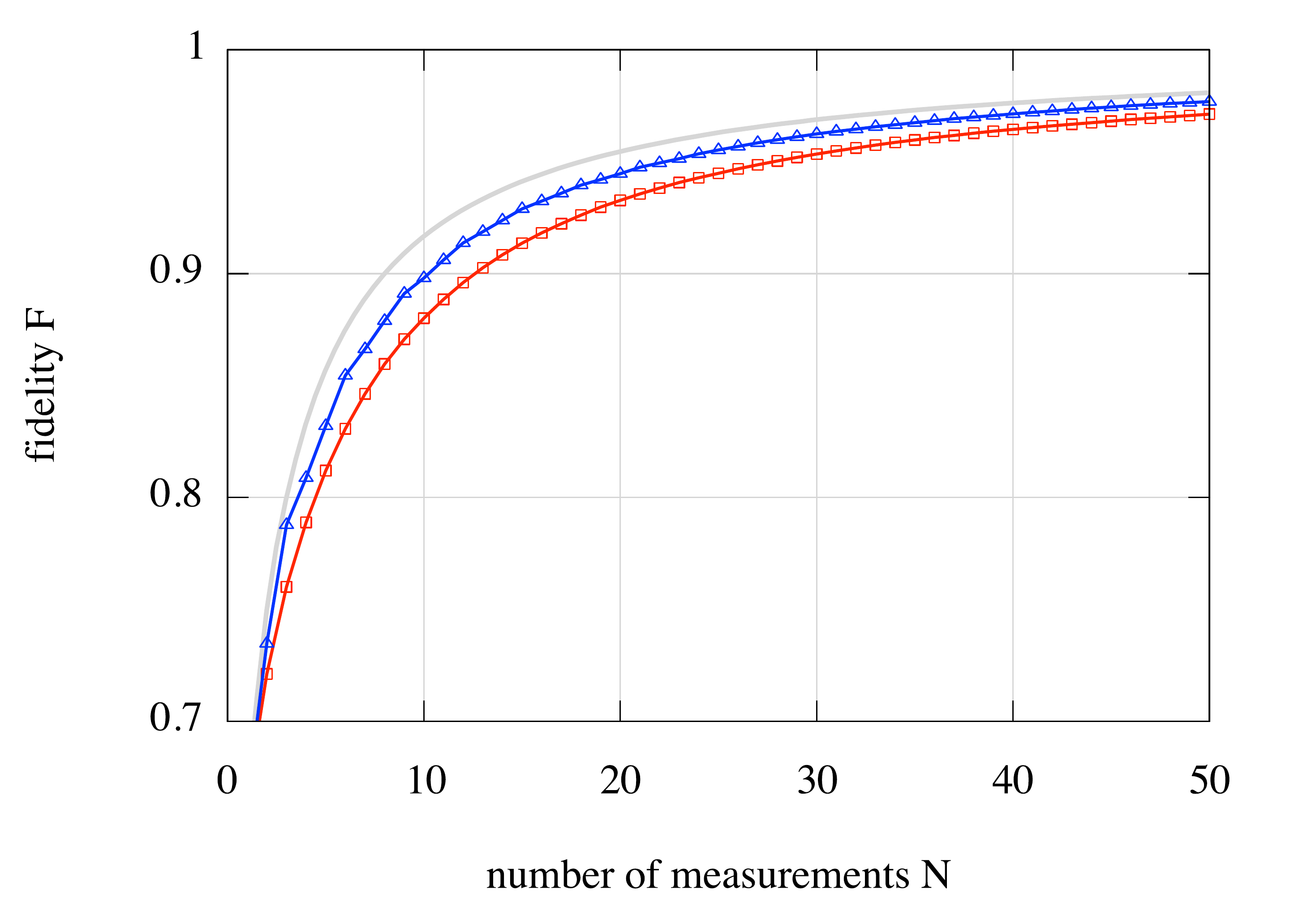}}

\subfloat[fidelity difference]{\includegraphics[width=0.5\textwidth]{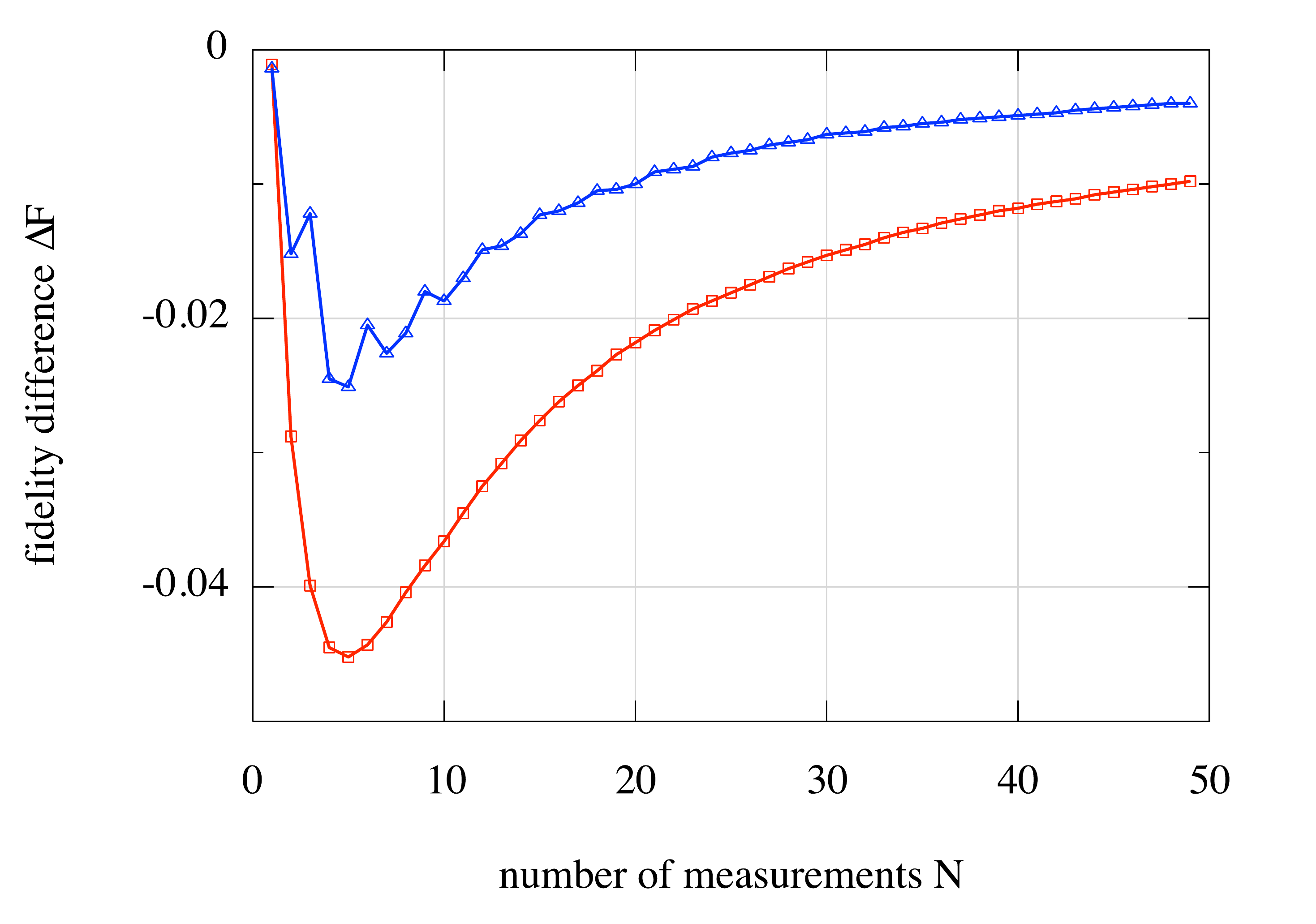}}

\caption{Least biasedness adaption for qubits. The blue triangles denote least
biasedness adaption, using equation~(\ref{eq:BiasCont}), compared
to random measurements (red squares) and the theoretical maximum (grey line)\label{fig:LBA2}.
In (a) the absolute estimation fidelities are shown, whereas (b) depicts the difference $\Delta F=F-F_{\mathrm{opt}}$ between the simulation fidelities $F$ and the optimal fidelity (\ref{eq:Fopt2}).}
\end{figure}

\begin{figure*}[t]
\subfloat[$d=4$]{\includegraphics[width=0.5\textwidth]{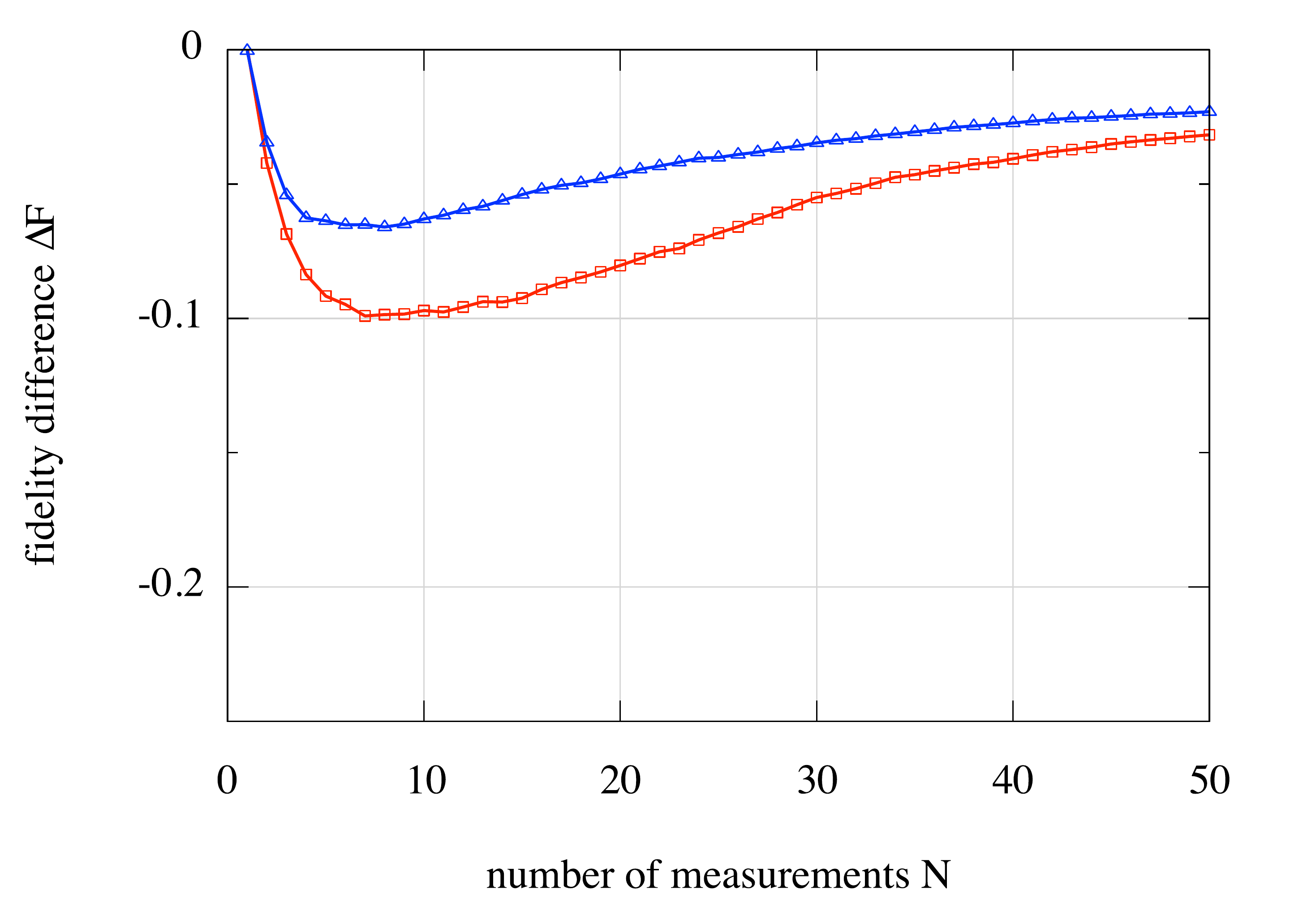}}
\subfloat[$d=6$]{\includegraphics[width=0.5\textwidth]{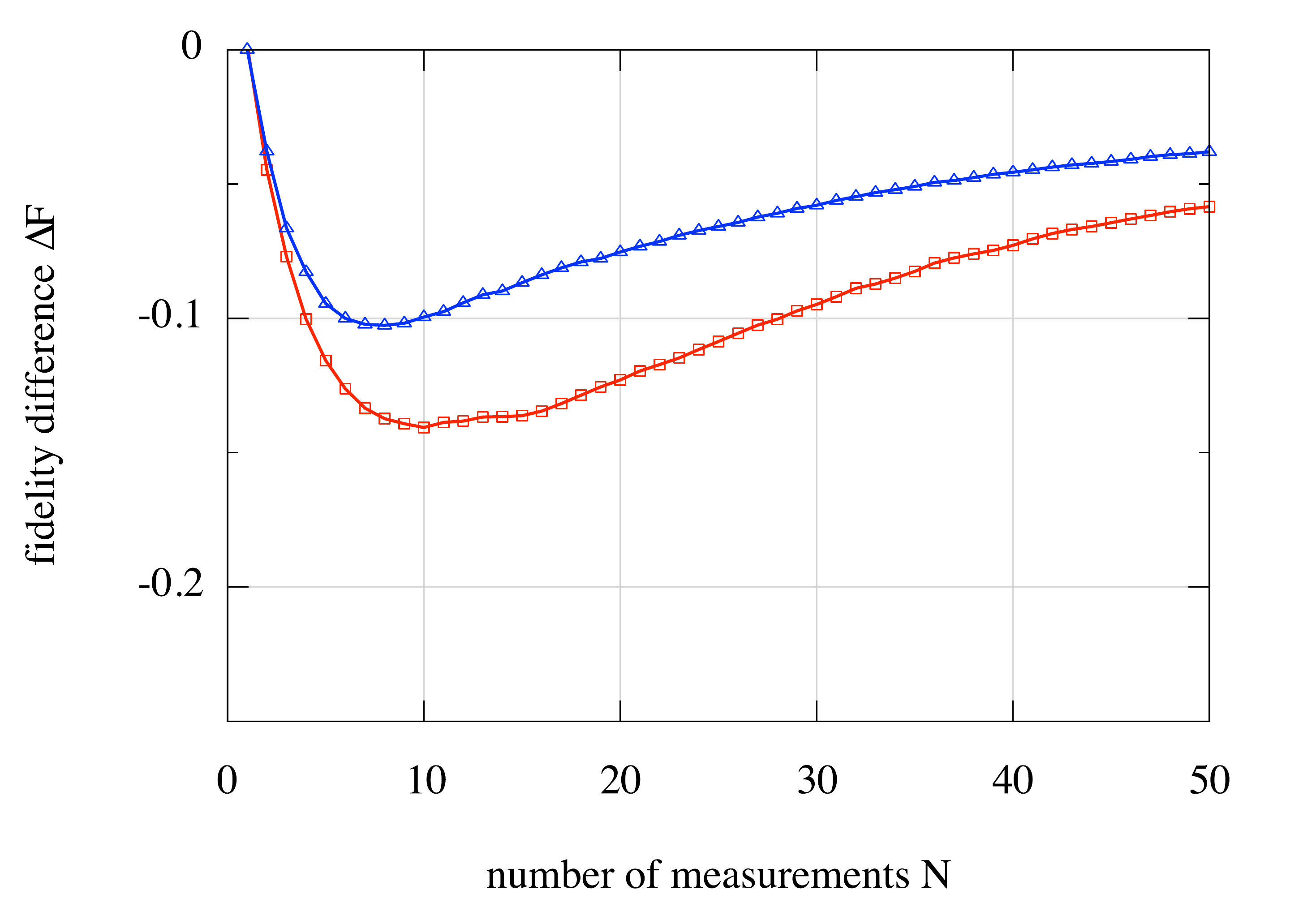}}\\
\subfloat[$d=8$]{\includegraphics[width=0.5\textwidth]{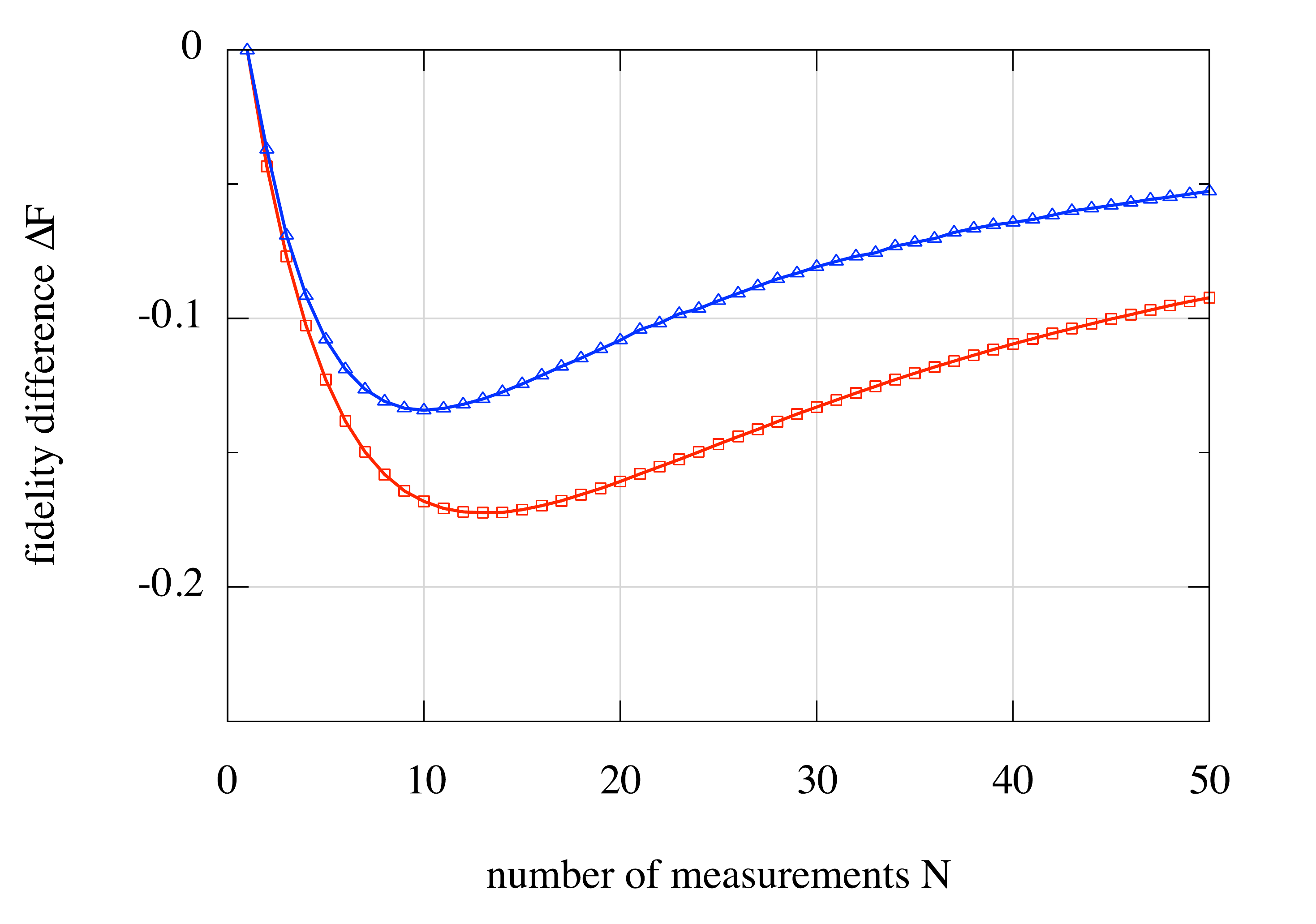}}
\subfloat[$d=13$]{\includegraphics[width=0.5\textwidth]{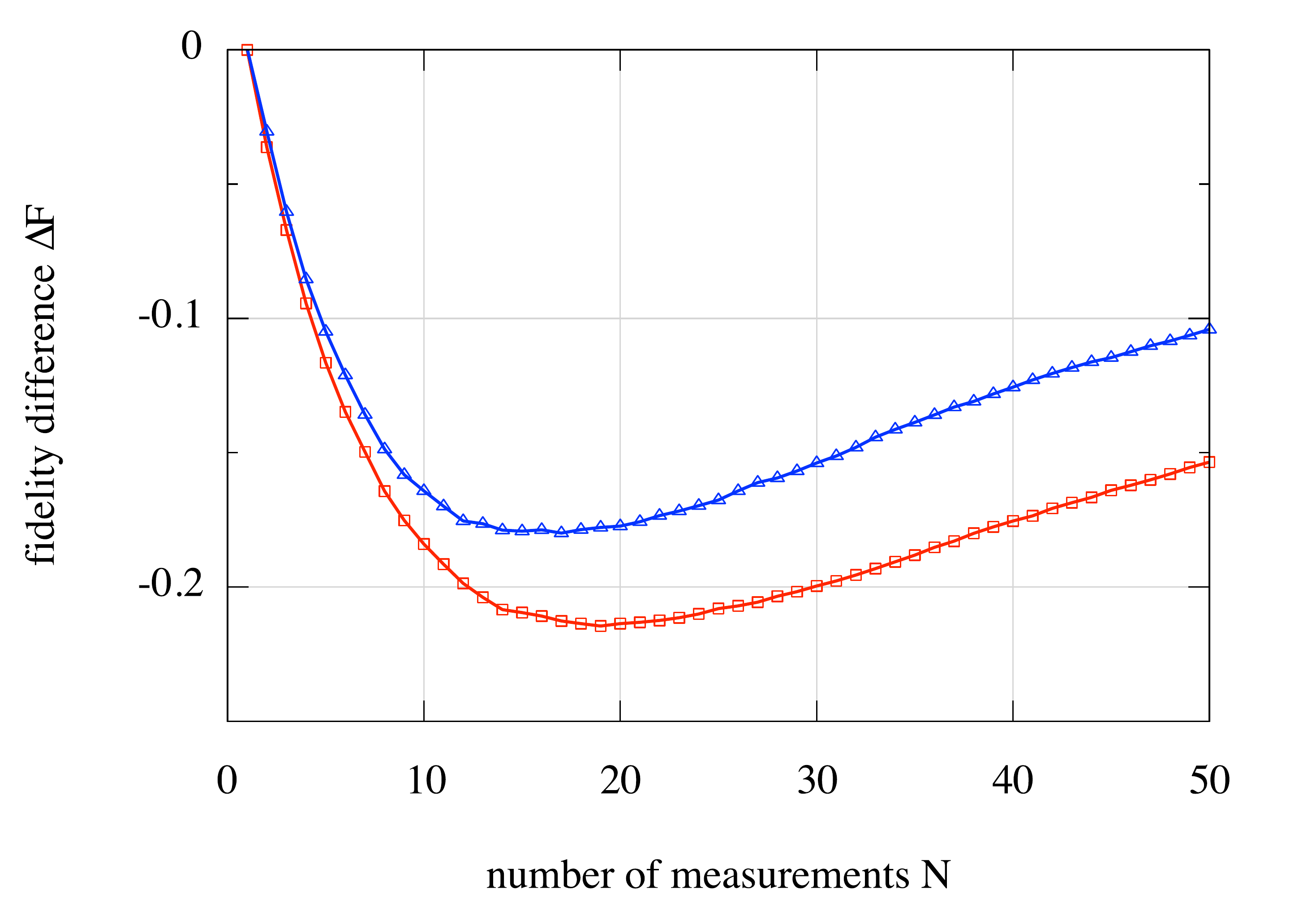}}

\caption{Least biasedness adaption for different dimensions. The fidelity difference
$\Delta F=F-F_{\mathrm{opt}}$ to the optimal fidelity (\ref{eq:Foptd})
is shown for four exemplary dimensions $d$. The blue triangles denote least biasedness adaption compared to random measurements
(red squares)\label{fig:LBAd}.}
\end{figure*}

The described method can be used for adaptive estimation of qubits.
The result of the corresponding simulations is depicted in figure~\ref{fig:LBA2}.

The resulting estimation fidelities are close to the theoretical maximum
\begin{equation}
F_{\mathrm{opt}}=\tfrac{N+1}{N+2} \label{eq:Fopt2}
\end{equation}
of collective measurements \cite{MP95,DBE98}.
Furthermore they are in good agreement with results reported in \cite{FKF00},
although here much simpler methods are used for adaption and estimation.

\subsection{Qudits}

The real test for the presented methods is their application to higher
dimensional states. In figure~\ref{fig:LBAd} we show the estimation
fidelities for states of dimension four, six, eight and 13. 

In all these examples, 
there is a gain in estimation fidelities due to adaption
compared to the non-adaptive procedure. On the other hand, in contrast
to the qubit case, there is a significant gap between adaptive single
measurements and theoretical maximal fidelity
\begin{equation}
F_{\mathrm{opt}}=\tfrac{N+1}{N+d} \label{eq:Foptd}
\end{equation}
in $d$ dimensions \cite{BM99}.

The case $d=6$ is of particular interest. 
Since $6$ is not a prime power, 
there is no known way to construct the seven MUBs necessary
for a Wootters-Fields measurement strategy \cite{WF89}. In fact there are only
three MUBs known for $d=6$ \cite{BW10}. But least bias adaption
manages to find six measurement bases unbiased with respect to the
measured vectors. Again, this does not mean that the measurement bases
are mutually unbiased, but that they have an overlap of $\frac{1}{6}$
only with the basis vectors corresponding to the really observed measurement
outcomes. Nonetheless the adaption fidelities are the same that would
be achieved with six fixed MUBs. For the seventh and all following measurements
the adapted bases are no longer {}``unbiased'' but have only least
bias with respect to the previously measured states.

\section{Conclusions}

We have shown that simple adaptive measurement methods can be successfully
applied to higher dimensional quantum states. We presented a method
for estimating and adapting in higher dimensions which reconciles the
concepts of adaption by information entropy maximisation and mutually
unbiased bases. 

The simulated estimation fidelities using these methods, 
although smaller than the theoretical maximum, 
show a clear fidelity gain
of up to $5\%$ compared to non-adaptive methods. This is independent
of the dimension and therefore also works for dimensions for which
no full set of mutually unbiased bases is known.\\

\bibliographystyle{apsrev4-1}
\bibliography{bibliography}

\end{document}